\documentclass[a4paper]{cas-sc}

\usepackage[numbers,compress]{natbib}
\setcitestyle{numbers,square,comma,sort&compress}
\usepackage{xcolor}
\usepackage{amsmath}
\usepackage{hyperref}
\usepackage[normalem]{ulem}
\def\tsc#1{\csdef{#1}{\textsc{\lowercase{#1}}\xspace}}
\tsc{WGM}
\tsc{QE}
\tsc{EP}
\tsc{PMS}
\tsc{BEC}
\tsc{DE}

\usepackage{amsmath}
\usepackage{graphicx}
\usepackage{booktabs}

\newcommand{\stirling}[2]{\genfrac{\{}{\}}{0pt}{}{#1}{#2}}
\begin{document}
\let\WriteBookmarks\relax
\def\floatpagepagefraction{1}
\def\textpagefraction{.001}

\shorttitle{Network exploration by random walks: A large deviation perspective}
\shortauthors{Upadhyay et al.}

\title [mode = title]{Network exploration by random walks: A large deviation perspective}

\author[1]{Sarvesh K. Upadhyay}[orcid=0009-0002-8947-5311]
\fnmark[1]
\ead{sarveshupadhyay@bhu.ac.in}

\author[2]{Trifce Sandev}[orcid=0000-0001-9120-3847]
\ead{trifce.sandev@manu.edu.mk}

\author[1]{Sanjay Kumar}
\ead{ksanjay@bhu.ac.in}

\author[3]{R. K. Singh}[orcid=0000-0002-2816-2218]
\cormark[1]
\ead{rksinghmp@gmail.com}

\affiliation[1]{organization={Department of Physics, Banaras Hindu University},
               city={Varanasi},
               postcode={221005},
               country={India}}

\affiliation[2]{organization={Macedonian Academy of Sciences and Arts},
               city={Skopje},
               country={Macedonia}}

\affiliation[2]{organization={Ss. Cyril and Methodius University},
               city={Skopje},
               country={Macedonia}}

\affiliation[2]{organization={Korea University},
               city={Seoul},
               country={Republic of Korea}}

\affiliation[3]{organization={University of Louisiana at Lafayette},
               city={Lafayette},
               state={LA},
               postcode={70503},
               country={USA}}

\cortext[1]{Corresponding author}

\begin{abstract}
We study exploration properties of a random walk on a network. For a fully connected network we find that the problem can be
mapped to the well known coupon collector problem, thus allowing us to estimate form of $P(S,t)$: the distribution of number of
distinct nodes $S$ visited by the random walk upto time $t$. From a practical point of view, however, both the fully connected
network and hops taking place after fixed intervals are an idealization. We solve this problem by introducing the formalism of
continuous time random walks wherein the random walk spends a random amount of time a node before hopping to its neighboring
node. The formalism allows us to study the large deviation limit of $P(S,t)$ under very mild conditions that the distribution of waiting times  $\psi(\tau)$ exhibits analyticity at small times. Furthermore, we find that at small times, the properties of $P(S,t)$ are largely independent of the network topology, and are governed solely by the waiting time characteristics.
\end{abstract}


\maketitle

\section{Introduction}
 The study of random walks on complex networks has emerged as a fundamental framework for understanding a wide range of dynamical processes \cite{pastor2015,bisnik2007,perra2012,sarkar2011,lopez2012,gkantsidis2004,lu2016}. Applications span diverse areas, including community detection \cite{fortunato2010,ballal2022,carletti2021,de2018}, ranking algorithms such as PageRank \cite{fortunato2007,agarwal2007}, transport phenomena \cite{barrat2008}, and the modeling of bursty spreading patterns in diseases and rumors \cite{khrennikov2020,draief2011,iannelli2017}. A fundamental quantity characterizing the exploration dynamics of a random walk is the number of distinct nodes (or sites) visited within a given time. This observable serves as a key measure of the efficiency of stochastic search and transport processes, ranging from animal foraging \cite{GORDON1995} to molecular trapping and diffusion in complex media \cite{Hollander1994,Havlin1984}. 
While typical behavior has been extensively studied in various contexts 
\cite{Gall1991,Gillis1970,Biroli_2022,Regnier2022,noh2004,MASUDA2017,hughes1995}, for example, the mean number of distinct nodes 
visited in $n$ steps has been analyzed on lattices \cite{Vineyard_1963}, and further extended to a collection of independent random 
walks, at both small and large times
\cite{Larralde1_1992,Yuste_2000}. On more complex structures, record statistics and covering-time distributions for random walks on 
fully connected lattices have been shown to exhibit both typical and extreme fluctuation regimes \cite{Turban_2015}. 
Collective scaling effects have been reported for multiple random walks on scale-free networks \cite{Aanjaneya_2019}, generating-
function approaches have been developed for random networks \cite{Bacco_2015}, and related quantities such as the mean number of 
common nodes \cite{Majumdar_2012}, and the time required by a random walk to find a previously not visited node conditioned on
information that a part of the network has already been explored \cite{Regnier2023}, have provided different measures for the 
exploration. 
Notwithstanding, a majority of these studies have focused on mean properties, and much less is known about rare realizations that 
significantly deviate from the average. Such events, where exploration occurs 
much faster than typical by visiting a disproportionately large number of nodes in short times, find striking analogies across real 
world systems. Malware outbreaks such as the Code-Red worm can spread rapidly across computer networks \cite{moore2002,weaver2003}, 
epidemiology provides striking examples of  trigger exceptionally rapid and widespread transmission, leading to  epidemic explosions. 
Such events highlight infectious diseases to propagate at unexpectedly fast rates \cite{hethcote2000,pastor2001}. In biomedical 
contexts, cancer metastasis exemplifies the sudden colonization of distant organs by malignant cells \cite{steeg2016,fidler2003}, and 
in ecology, invasive species destabilize ecosystems through large scale habitat colonization \cite{pimentel2005,mack2000}. Likewise, 
in social systems, bursts of misinformation can trigger rumor cascades that reach massive audiences within short time scales \cite{Vosoughi2018}.

These phenomena raise a natural question: How does the number of distinct nodes $S$ visited evolve over time, and can we predict the likelihood of extreme, fast-spreading trajectories that drive catastrophic events in networks? To address this, we develop a framework that  provides us with the full distribution $P_n(S)$ of distinct nodes visited on a fully connected network for  a random
walk (RW) in $n$ jumps. We extend this result to estimate
the distribution of the number of distinct nodes $P(S,t)$ explored by a continuous time random walk (CTRW) upto 
time $t$. It is to be noted here that going from RW to a CTRW is a nontrivial extension in the sense that we are not only
generalizing to random waiting times, but we also go beyond the realm of fully connected networks and study practically
relevant cases like transport on heterogeneous networks.
Our results capture not only typical exploration behavior but also rare, high-impact events. All analytical predictions are validated through numerical simulations, showing excellent agreement.
\section{Network exploration by a RW} 
\begin{figure}
    \centering
    \includegraphics[width=0.494\linewidth]{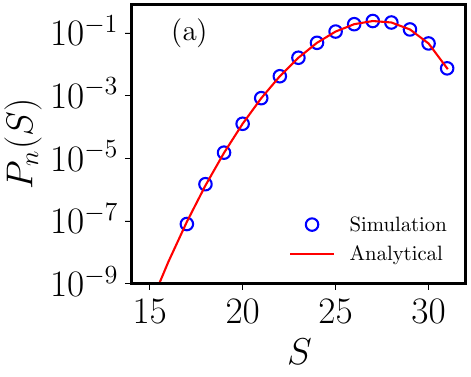}
\includegraphics[width=0.494\linewidth]{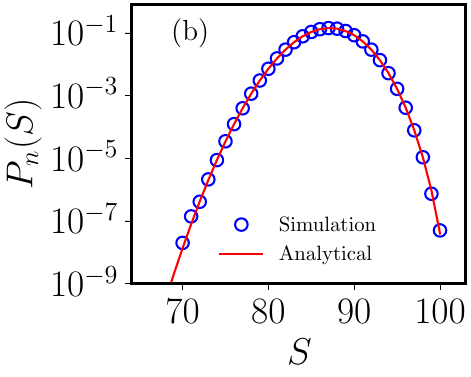}
    \caption{Distribution of distinct visited nodes $P(S,t)$ for a discrete time random walk on a complete network ($N=100$). 
    (a) $t=30$. (b) $t=200$.}
    \label{fig:Complete_net_s(t)_discrete}
\end{figure}

Consider a discrete time random walk (RW) exploring
an undirected network consisting of $N$ nodes. In this discrete time formulation, each step of the RW corresponds to a single instantaneous transition between two connected nodes. Therefore, time is naturally measured in terms of the number of jumps, allowing us to identify $n = t$. Let $P_{ij}(n)$ denote the 
probability that RW, starting from node $i$ at time $n = 0$, 
is found at node $j$ after $n$  jumps. 
The evolution of this probability is governed by the master equation \cite{noh2004}:

\begin{equation}
P_{ij}(n + 1) = \sum_{k} \frac{A_{kj}}{k_k} \, P_{ik}(n),
\label{eq:master_discrete}
\end{equation}
where $A_{kj}$ represents the adjacency matrix of the network 
($A_{kj} = 1$ if nodes $k$ and $j$ are connected, and $0$ otherwise), 
and $k_k$ denotes the degree of node $k$. 
 Let the random walk visit $S$ distinct nodes after performing $n$ jumps. Since each jump can introduce at most one newly visited node, the number of distinct nodes visited satisfies $S \le n + 1$. Moreover, because the network contains only $N$ nodes in total, we also have $S \le N$. Hence, the number of distinct nodes is bounded as $S \le \min\{n + 1,\, N\}.$
At the initial time, $S(0) = 1$, meaning that the  RW starts from a single node and has $N - 1$ remaining nodes to explore. After $n$ jumps, the  RW has discovered $S - 1$ new nodes. For a fully-connected network: $A_{kj} = 1-\delta_{ij}$, with $\delta_{ij} = 1$
whenever $i = j$ and $0$ otherwise, is Kronecker's delta. 
 As a result, each  jump corresponds to a uniform random sampling among the $N - 1$ remaining nodes. The spatial homogeneity implies that the order in which distinct nodes are discovered depends solely on the sequence of random selections, and not on any geometric constraints. Consequently, the exploration process can be described exactly by combinatorial counting.
Each discovery of a new node defines a renewal event, partitioning the trajectory into $S - 1$ intervals corresponding to the steps between successive discoveries. The number of ways to partition $n$ jumps into $S - 1$ nonempty subsets is given by the Stirling numbers of the second kind  $\Big\{{n \atop S-1}\Big\}$ \cite{Graham1994}.  Thus, the total number of distinct sequences for discovering $S-1$ new nodes is $(S - 1)! \Big\{{n \atop S - 1}\Big\}$.
Furthermore, the $S - 1$ distinct nodes themselves can be chosen from the remaining $N - 1$ nodes in $\binom{N - 1}{S - 1}$ possible ways. As a result, the total number of exploration trajectories that result in exactly $S$ distinct visited nodes after $n$ jumps is 
\begin{equation}
    f_n(S) = \binom{N - 1}{S - 1} (S - 1)! \Big\{{n \atop S - 1}\Big\}
    \label{f_n(S)}
\end{equation}
. This normalization can be verified using the Stirling identity \cite{BaguiMehra2024},
\[
x^n = \sum_{k = 0}^{n} \Big\{{n \atop k}\Big\} x^{\underline{k}},
\quad \text{with} \quad x^{\underline{k}} = k!\binom{x}{k},
\]
which, for $x = N - 1$, gives
\begin{equation}
    \sum_{S = 1}^{\min(n + 1, N)} \Big\{{n \atop S - 1}\Big\} (S - 1)! \binom{N - 1}{S - 1} = (N-1)^n
    \label{(N-1)^n}.
\end{equation}
 This
implies that the distribution for the number of distinct nodes visited after $n$ jumps is:
\begin{equation}
\label{eq:S_distribution}
P_n(S) =
\binom{N - 1}{S - 1}
\frac{(S - 1)!\, \Big\{{n \atop S - 1}\Big\}}{(N - 1)^n},~1 \leq S \leq N.
\end{equation}
\noindent
 It is to be noted here that this combinatorial approach works because of the network’s complete symmetry and the  fact
that each jump is statistically independent, except that already visited nodes cannot count as new. 
Summarily, the exploration problem  is equivalent to the classical coupon collector problem (CCP), where each node represents a distinct coupon type being collected over time \cite{Tishby_2022}. The analytical prediction in Eq.~(\ref{eq:S_distribution}) shows excellent agreement with simulations,  see Fig.~\ref{fig:Complete_net_s(t)_discrete}.

What is the ultimate limit of exploration? Once  the RW has visited $S$ distinct nodes, a natural question arises: how long does it take to visit all nodes?
This question leads us to the concept of the cover time, defined as
$T_{\mathrm{cov}} = \inf \{ t \;|\; S(t) = N \}$,
 and represents the minimum time required for the  RW to visit every node in the network at least once.  
In other words, while $S(n)$ quantifies the progressive growth of the explored region, $T_{\mathrm{cov}}$ captures the extreme tail event where the entire network has been covered. Formally, $T_{\mathrm{cov}}$ is a stopping time associated with the process $\{S(n)\}$. For the completely connected network, the cover time problem is identical to the  CCP \cite{Aldous1983,Wilf01101989,feller1968,Adler2003}.  
In the CCP, each “throw” corresponds to a random selection of one among $N$ coupon types, and the goal is to determine  $n_{\mathrm{cov}}$, the number of throws  needed to collect all types at least once.  
The expected number of throws required to complete the collection is $\langle n_{\mathrm{cov}} \rangle = N H_N$, where $H_N = \sum_{k=1}^N \tfrac{1}{k}$ is the $N$-th harmonic number.  
For a complete network \emph{without} self-loops, the walk cannot remain on the same node between successive jumps, effectively reducing the number of available choices at each step from $N$ to $N-1$.  
Consequently, the connection between the CCP and the exploration of a fully-connected network
by an RW implies that the mean cover time  of the RW is
\begin{equation}
\langle T_{\mathrm{cov}} \rangle = (N - 1) H_{N - 1},
\end{equation}
\noindent
 and connects the microscopic statistics of partial exploration to the macroscopic timescale of full coverage. 
\section{Network exploration by a CTRW} The jump description of an RW discussed above captures only the sequence of visited nodes, neglecting the physical time spent moving between nodes. To connect exploration dynamics more closely to real transport processes such as diffusion~\cite{pastor2015}, communication~\cite{Boguna2009}, and energy or transport phenomena on networks~\cite{barrat2008}, we extend the formulation to a continuous time setting. In the  \textit{continuous time} formalism, the RW moves across the  network structure but now experiences random waiting times between successive jumps,  making it a CTRW. This is relevant across systems that exhibit a non-degenerate residence time distribution, for example, in particle transport through porous media and other disordered environments~\cite{de1986hydrodynamic,sahimi2012dispersion,varloteaux2013pore}. 

 Network exploration  by a CTRW involves two independent stochastic 
components:
$(i)$ the sequence of visited nodes, governed by the  jump process analyzed earlier, and
$(ii)$ the timing of jumps, determined by waiting times $\tau$ which are independent and identically distributed
(IID) random variables following the distribution $\psi(\tau)$.

 Since the choice of the next node is independent of how long the  CTRW waits before moving, these two sources of randomness can be treated separately and then recombined. Let $Q_t(n)$ denote the 
probability that exactly $n$ jumps have occurred up to time $t$. In addition, the number of jumps 
$n$ depends solely on the temporal process, and not on the topology or the sequence of visited nodes. On the other hand, $P_n(S)$, the probability of visiting $S$ distinct node in $n$ jumps, depends only on the  sequence of jumps.  As a result, distribution of $S$ at time $t$,  that is, $P(S,t)$, can be expressed through the subordination 
equation:
\begin{equation}
P(S,t)=\sum_{n=0}^\infty P_n(S)\,Q_t(n),
\label{eq:subordination_explicit}
\end{equation}
 and holds whenever node selection and waiting times are independent, and 
 \begin{equation}
     Q_t(n) = \int^t_0 d\tau~\psi_n(\tau)\Psi(t-\tau)
 \end{equation}
\cite{montroll1965,METZLER2001}, where $\psi_n(\tau)$ is the $n$-fold convolution of the waiting-time distribution $\psi(\tau)$ and $\Psi(t) = 1-\int^t_0 du~\psi(u)$.
 Now, for a general network topology and an arbitrary waiting time distribution, estimating
$P(S,t)$ from the subordination equation in (\ref{eq:subordination_explicit}) is a hard problem. However,
for the special case of a fully-connected network, $P_n(S)$ is described by Eq.~(\ref{eq:S_distribution}).
For further simplicity, let us assume that the distribution of waiting times is exponential, that is,
$\psi(\tau) = \lambda e^{-\lambda\tau}$, which implies that $Q_t(n)=(\lambda t)^ne^{-\lambda t}/n!$
\cite{feller1968,Barkai2020}. As a result, for a CTRW jumping from one node to another at a constant
rate $\lambda$ on a fully-connected network, the distribution of the number of distinct nodes is:

\begin{equation}
\label{eq:poisson}
P(S,t)=\sum_{n=0}^\infty 
\binom{N-1}{S-1}\,
\frac{(S-1)!\,\stirling{n}{S-1}}{(N-1)^n}\,
e^{-\lambda t}\frac{(\lambda t)^n}{n!},
\end{equation}
which is simply a Poisson mixture of the $n$-step distributions.
 In addition, $P(S,0)=\delta_{S,1}$,
as initially only the starting node is occupied.
\begin{figure}
    \centering
    \includegraphics[width=0.49\linewidth]{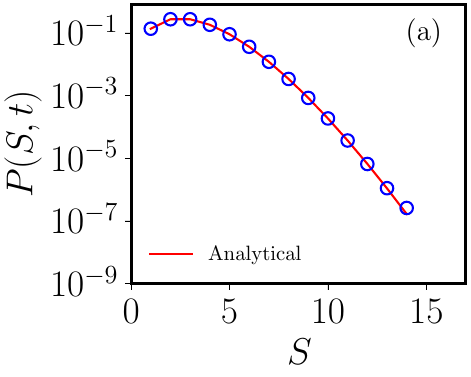}
    \includegraphics[width=0.49\linewidth]{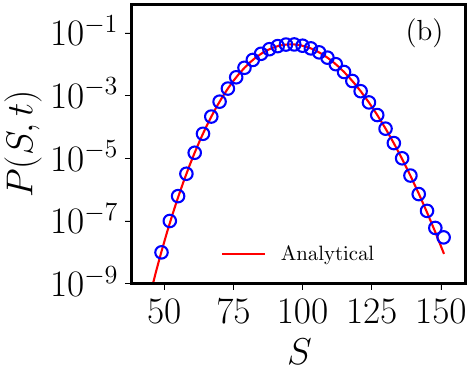}
    \caption{Distribution of distinct visited nodes $P(S,t)$ for a CTRW on a complete network ($N=1000$, $\lambda=2$)  at small ((a) $t = 1$) and large ((b) $t = 50$) times and following Eq.~(\ref{eq:poisson}).
    }
    \label{fig:Complete_net_s(t)}
\end{figure}
 Now, at small times, the mean number of distinct nodes explored by a CTRW with constant
jump rates is: $\langle S(t) \rangle \stackrel{\text{small}~t}{\sim} 1+\lambda t$ \cite{Upadhyay2025} {see End matter
for details).
This implies that the distribution $P(S,t)$ described in Eq.~(\ref{eq:poisson})
 captures both the typical and rare fluctuations in the exploration process,
 see Fig.~\ref{fig:Complete_net_s(t)}.

\begin{figure}
    \centering
    \includegraphics[width=0.60\linewidth]{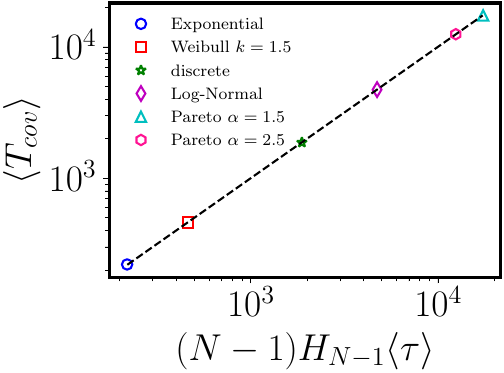}
   \caption{Mean cover time $\langle T_{\mathrm{cov}} \rangle $ for various waiting-time distributions on networks with 
$N=50,100,300,600,800,1000$ nodes. The distributions  are:  
$\psi(\tau) = \lambda e^{-\lambda \tau}$ with $\lambda=1$ (exponential);  
$\psi(\tau) = \tfrac{k}{\lambda}\left(\tfrac{\tau}{\lambda}\right)^{k-1} e^{-(\tau/\lambda)^k}$ 
with $\lambda=1,~k=1.5$ (Weibull);  
$\psi(\tau) = \tfrac{1}{\tau \sigma \sqrt{2\pi}} 
\exp\!\left[-\tfrac{(\ln \tau - \mu)^2}{2\sigma^2}\right]$ 
with $\sigma=0.5,~\mu=0$ (log-normal);  
$\psi(\tau) = \alpha \tau_m^{\alpha} \tau^{-1-\alpha}$ 
with $\tau_m=1$ and different $\alpha$ values (Pareto).  
The black dashed line follows  Eq.~(\ref{eq:mean_cover_time}).}
    \label{fig:Mean_cover_time}
\end{figure}

 Extending the properties of a RW, it is evident that for a CTRW with IID waiting times $\{\tau_m\}_{m\ge1}$ , the total cover time is given by $T_{\mathrm{cov}} = \sum_{m=1}^{n_{\mathrm{cov}}} \tau_m$. 
 Let us further assume that the distribution of waiting times $\psi(\tau)$ possesses a finite
mean $\langle \tau \rangle$. Then, $\langle T_{\mathrm{cov}} \rangle = \langle \sum_{m=1}^{n_{\mathrm{cov}}} \tau_m \rangle = \langle n_{\mathrm{cov}} \rangle \langle \tau \rangle$,
since the waiting times are independent of the trajectory. 
Using $\langle n_{\mathrm{cov}} \rangle = (N-1)H_{N-1}$ for  a RW on a 
fully-connected network, we obtain
\begin{equation}
    \langle T_{\mathrm{cov}} \rangle = (N-1) H_{N-1} \langle \tau \rangle
    \label{eq:mean_cover_time},
\end{equation}
which connects the mean cover time in continuous time to its discrete-time counterpart through the average waiting time $\langle \tau \rangle$. We study in Fig.~\ref{fig:Mean_cover_time} the $\langle
T_{\mathrm{cov}} \rangle$ for a RW and a CTRWs with various $\psi(\tau)$ exploring a fully-connected
network, and find that both the RW and the CTRW with a finite $\langle \tau \rangle$ explore the network
in an analogous manner. This is not surprising in light of the fact that a RW is behaves similar to
a CTRW with $\langle \tau \rangle = 1$, corroborating the assertion earlier put forward in 
Ref.~\cite{Upadhyay2025}.
\begin{figure}
\centering
\includegraphics[width=0.49\linewidth]{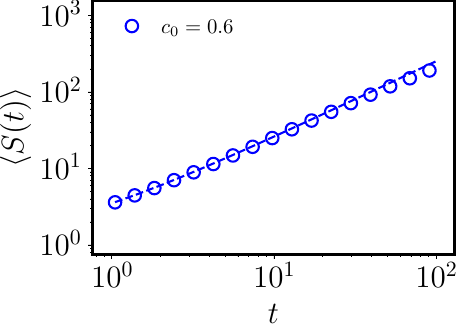}  
\includegraphics[width=0.49\linewidth]{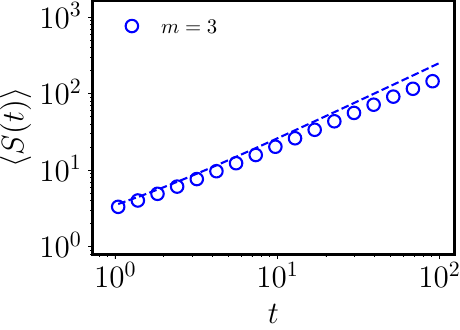}
\caption{
Small-time behavior of the mean number of distinct nodes visited up to time $t$ for a CTRW on an Erd\H{o}s–R\'{e}nyi
(ER) network with 
jump rate $r$: (left panel) $c_0 = 0.6,~ r = 2.5,~ N = 600$. Blue circles denote numerical estimates and black dashed 
lines follow: $\langle S(t) \rangle \sim 1 + \lambda t$, for  $t$ small. The right panel corresponds to the Barab\'{a}si-
Albert (BA) network
with parameters: $m = 3,~ r = 2.5,~ N = 600$, where $m$ is the number of edges that each newly added node forms
in the BA model.
\label{fig4}
}
\end{figure}

\section{Rare events and the large deviation regime}
Among various network topologies, the fully-connected network 
is unique in allowing an exact closed-form expression for the distribution of 
distinct nodes visited, $P_n(S)$, given by Eq.~\eqref{eq:S_distribution}. 
This exact solvability arises from the absence of spatial heterogeneity, since 
each step corresponds to a uniform random sampling over all nodes, effectively 
reducing the problem to the CCP. 
For other network structures, such as rings, lattices, trees, or sparse random graphs, 
the exploration dynamics become non-Markovian in $S$, as the probability of 
discovering a new node depends  strongly on the local connectivity and geometry of the 
already explored region. Consequently, deriving a closed form expression for
$P_n(S)$  becomes an extremely difficult task for RWs exploring these networks. 
 Nevertheless, the subordination relation in 
Eq.~\eqref{eq:subordination_explicit} remains valid for random walks on any 
network. At this point, we would also like to emphasize on the reason of explicitly focusing on small time behavior.
This is because at intermediate
to large times the distribution $P(S,t)$ is described by the central limit theorem, and is
a Gaussian centered around its mean. While it is also true that there exist rare fluctuations even at large times,
because at large times finding a new previously not visited node becomes increasingly
difficult. However, this is not a very useful regime in the sense that most
of the network has been explored, and the damage is already done, for example, if we are talking
in terms of a virus affecting the computer network. For this reason, we focus on the regime where the number of 
renewals is still small compared to the network size, that is, $1 \ll n \ll N$,
where the distribution becomes sharply peaked, 
so that $P_n(S) \approx \delta_{n,S-1}$.
Physically, such a scenario arises when the network exploration by a CTRW is still in
its infancy, that is, $t$ is small and every time the CTRW jumps to a node,
the chances of it being a new, previously unexplored
node is much higher compared to revisiting a previously visited node. This short-time behavior is fully consistent with our recent work \cite{Upadhyay2025}. At short times, the walker makes only a few steps and explores just the immediate neighborhood of its starting node, so the dynamics depend only on local connectivity rather than global structure. In Fig \ref{fig4}  we show the numerical simulations on both Erd\H{o}s--R\'enyi (ER) and Barab\'asi--Albert (BA)
networks confirm that the early-time growth of
$\langle S(t)\rangle$ follows $\langle S(t)\rangle \sim 1+\lambda t$, which is similar across different network types.
 This corroborates our assertion that at small times the dynamics depends only on the local structure.
Furthermore, the linear growth of $\langle S(t) \rangle$ implies that the distribution of number of distinct nodes
visited in $n$ jumps: $P_n(S) \approx \delta_{n,S-1}$. As a result,
\begin{figure}
    \centering
    \includegraphics[width=0.494\linewidth]{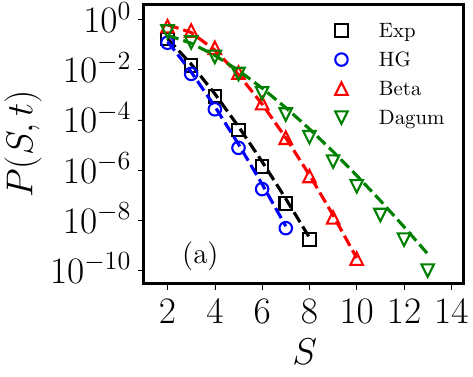}
    \includegraphics[width=0.494\linewidth]{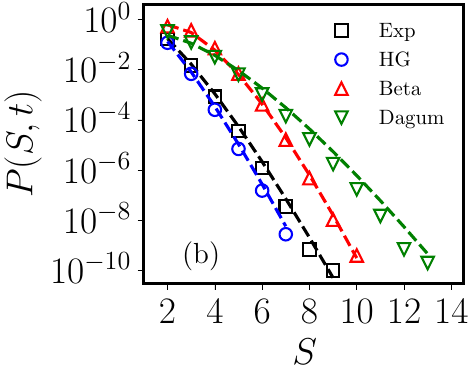}
   \caption{Probability distribution of the number of distinct nodes visited $P(S,t)$,
   by a CTRW on (a) an Erd\H{o}s–R\'{e}nyi (ER) network with $N=1000$ and connection probability $c=0.01$ (sparse  and homogeneous), and (b) a Barabási–Albert (BA) network with $N=1000$ and $m=4$ ( heterogeneous). 
Symbols represent CTRW simulations, and  dashed line follows  Eq.~\ref{eq:large_dev_PSt}.
  Different waiting time distributions
are: exponential (exp): $\psi(\tau)=\lambda e^{-\lambda\tau}$ with $\lambda=2$ and $t=0.1$, half-Gaussian (HG): $\psi(\tau)=\dfrac{2}{5\pi}e^{-\tau^{2}/(25\pi)}$ with $t=1$, 
beta:  $\psi(\tau)=6\tau(1-\tau)$ for $0\le\tau\le1$ with $t=1$, and  Dagum: $\psi(\tau)=\dfrac{1}{(1+\tau)^{2}}$ with $t=1$.}
    \label{fig:poission_shifted}
\end{figure}
\begin{align}
P(S,t) \stackrel{\text{small}~t}{\sim} 
\sum_{n=0}^{\infty}\delta_{n,S-1}\, Q_t(n)
= Q_t(S-1),
\label{poisson_shifted}
\end{align}
indicating that the statistics of the number of distinct  nodes visited upto time $t$
directly mirror the statistics of the renewal (jump) process itself. 
As discussed in the \textit{Introduction}, such rare events correspond to  an atypically rapid
exploration.  Hence, if the short time behavior of the distribution $Q_t(n)$ is known, then
problem of estimating $P(S,t)$ at short times is solved. Now, a broad class of waiting time distributions
$\psi(\tau)$ exhibit an analytic structure for small waiting times, that is,

\begin{equation}
    \psi(\tau) \stackrel{\text{small}~\tau}{\sim}
C_A \tau^A + C_{A+1}\tau^{A+1} + \cdots
\label{exp_psi}
\end{equation}
where $A \ge 0$ is an integer,  and $C_A$ and $C_{A+1}$ depending on the parameters of the
distribution $\psi(\tau)$ \cite{Barkai2020}.  Under these very mild conditions, $Q_t(n)$ 
admits a large deviation form
\cite{Barkai2020} (see Appendix for details):
\begin{equation}
Q_t(n)
\stackrel{\text{large}~n/t}{\sim}
\frac{1}{\sqrt{2\pi\,n(A+1)}}\;
\exp\!\left[-n\,I\!\left(\frac{t}{n}\right)\right],
\end{equation}
with the rate function:
\begin{equation}
    I\!\left(\frac{t}{n}\right)
\sim (A + 1)
\!\left[
- \ln\!\left(
\frac{e\, [C_A \Gamma(A + 1)]^{1/(A + 1)}}{(A + 1)}\frac{t}{n}
\right)
- \frac{C_A + 1}{(A + 1)C_A}\,\frac{t}{n}
\right]
\end{equation}

It is to be noted here that this rate function describes the domain of a large
number of renewals and depends only on the small time properties of the waiting time distribution
$\psi(\tau)$. This means that the rate function is defined independent of whether the mean waiting time ($\langle\tau\rangle$) exists.
Using this result with $n = S-1$ we find for small $t$:
\begin{equation}
P(S,t) \sim \frac{1}{\sqrt{2\pi\,(S-1)(A+1)}}
\exp\!\left[-(S-1)\, I\!\left(\frac{t}{S-1}\right)\right].
\label{eq:large_dev_PSt}
\end{equation}
The argument $t/(S-1)$ represents the effective time per visit and governs the rarity of  network 
exploration  at small times, 
and capturing the explosive spreading phenomena such as super spreading events in epidemics, where a single individual drives an exceptionally rapid epidemic cascade. Another example is malware outbreaks that occur much faster than the typical dynamics.

The result in Eq.~(\ref{eq:large_dev_PSt}) is of fundamental importance, and provides the large-$S$ limit of $P(S,t)$ for small $t$. As discussed above, in such a scenario the
CTRW almost always arrives at a new, previously unexplored node, every time it makes a jump. This means that
the small-$t$ exploration properties of a CTRW do not depend on the network topology, whether the network
is homogeneous or heterogeneous (see Fig.~\ref{fig:poission_shifted}). The approximation 
$\delta_{n,S-1}$ is  nearly exact for the small time $t$  for complete graphs and dense ER networks, where simulations and theoretical prediction from Eq.~\ref{eq:large_dev_PSt} coincide trivially well; hence, we do not show those cases. Instead, Fig. ~\ref{fig:poission_shifted} highlights the more challenging regimes, sparse ER networks and BA networks with $m=4$
where connectivity is limited and degree heterogeneity is significant. Even here, the large-deviation prediction matches CTRW simulations closely. Furthermore, the large deviation
form of $P(S,t)$ in Eq.~(\ref{eq:large_dev_PSt}) implies that at small times, the exploration properties of
a CTRW are almost independent of whether or not $\langle\tau\rangle$ exists. This contrasts sharply with
the average exploration properties of a CTRW, whose temporal evolution of $\langle S(t) \rangle$
exhibits drastically different behaviors depending on the (non)-existence of $\langle\tau\rangle$
\cite{Upadhyay2025}.

\section{Discussion}
In this work, we  present an analytical framework for the exploration dynamics of random walks on networks. 
 We map the dynamics of random walks (RWs) to the coupon collector
problem,  and find an exact   expression of the distribution of  the number of distinct nodes $P_n(S)$, explored by the
RW on a fully-connected network in $n$ jumps. The mapping further allows us to write the exact form of the mean cover time $\langle T_{\text{cov}} \rangle$ taken by the RW to visit each node of the fully-connected network
at least once. We also generalize the study to include the practically relevant case of RWs with non-degenerate
distribution of waiting times, that is, continuous time RWs exploring a network. Under the simplifying assumption 
that waiting times are independent of the jump dynamics, we extend the concept of subordination to
exploration of a network by CTRWs. We find that the rare fluctuations in $P(S,t)$ for small $t$ exhibit
a large deviation form that is independent of the structural details of the network. 
Together, these results establish a complete analytical foundation for understanding exploration processes on undirected networks. Extending this framework beyond the fully-connected network remains a central challenge. Recently, a large-deviation approach was used to study distinct-site exploration behavior in the tails of the distribution for lattices with $d \ge 2$ \cite{smith_2025}.  In structured topologies such as rings, lattices, or sparse random graphs, exploration becomes inherently non-Markovian due to geometric constraints. Developing asymptotic and recursive approaches for such systems, and incorporating non-exponential waiting times ~\cite{METZLER2001}, would bridge bursty transport ~\cite{khrennikov2020,
draief2011,iannelli2017}, anomalous diffusion ~\cite{METZLER2001}, and rapid spreading phenomena ~\cite{hethcote2000,pastor2001} across diverse networks ~\cite{Holme2012}. 
This framework thus lays the groundwork for developing deep insights into the mechanisms governing network exploration, information flow, and epidemic propagation in complex systems~\cite{pastor2015,barrat2008}.

\section*{Acknowledgments}
SKU acknowledges support from the UGC Fellowship (Award Reference No.- 211610081005) and funding from SERB, New Delhi, and IoE(BHU), MoE, New Delhi, India  for computational resources. TS acknowledges financial support by the German Science Foundation (DFG, Grant number ME 1535/12-1) and by the Alliance of International Science Organizations (Project No. ANSO-CR-PP-2022-05). TS was also supported by the Alexander von Humboldt Foundation.
\printcredits
\bibliographystyle{unsrt}
\bibliography{cas-refs}

\section*{\centering Appendix}
\subsection*{Form of the rate function $I(t/n)$}Starting from $Q_t(n)$ given in Ref.~\cite{Barkai2020}:
\begin{equation}
Q_t(n)
\stackrel{\text{large}~n/t}{\sim}
\frac{\left[(C_A \Gamma(A+1))^{1/(A+1)}\, t\right]^{n(A+1)}}{\Gamma\!\left(n(A+1)+1\right)}
\exp\!\left(\frac{C_{A+1}}{C_A}\,t\right),
\label{eq:Qt_asym_before_stirling_n}
\end{equation}
define $z = n(A+1)$ and $B \equiv \big(C_A \Gamma(A+1)\big)^{1/(A+1)}$, and applying
Stirling's approximation: $\Gamma(z+1) \approx \sqrt{2\pi z}\; z^{\,z}\, e^{-z}$, we rewrite the
above equation as
\begin{align}
Q_t(n) \stackrel{\text{large}~n/t}{\sim}
\frac{1}{\sqrt{2\pi z}}
\exp\!\left[
z\ln(Bt) - z\ln z + z
+ \frac{C_{A+1}}{C_A}\,t
\right].
\label{eq:Qt_after_stirling_n}
\end{align}
Now, define the rate function associated with the large deviation form of $Q_t(n) \stackrel{\text{large}
~n/t}{\sim} \exp[-nI(t/n)]$, and comparing this definition to Eq.~(\ref{eq:Qt_after_stirling_n}) we
find the rate function
\begin{align}
I\!\left(\frac{t}{n}\right)
=
(A+1)
\Big[&
- \ln\!\left(
\frac{e\,(C_A\Gamma(A+1))^{1/(A+1)}\, t}{n(A+1)}
\right)\nonumber 
- \frac{C_{A+1}}{(A+1)C_A}\,\frac{t}{n}
\Big],
\label{eq:rate_fn_full_n}
\end{align}
which is used in the main text in Eq.~(\ref{eq:large_dev_PSt}) to describe the rare fluctuations
in the distribution of distinct nodes $P(S,t)$ visited by the CTRW upto time $t$.

\end{document}